# Relativistic velocity transformation as a genitor of transformation equations (relativistic dynamics)

**Bernhard Rothenstein Politehnica University Timisoara Physics Department Timisoara Romania**

*Abstract. The fundamental equations of relativistic dynamics are derived from a thought experiment and from the transformation of relativistic velocity avoiding collisions and conservation laws of momentum and energy.*

## 1. Introduction

Thought experiments are an antidote for counterintuitive effects associated with special relativity. Physicists conceive them in order to prove the reality of the time dilation and length contraction relativistic effects, invoking the principle of relativity. Thought experiments and the principle of relativity lead to the relativistic velocity transformation, without using the Lorentz-Einstein transformations.[1,2,3,4,5]

The purpose of our paper is to derive transformation equations for the physical quantities introduced in relativistic dynamics: mass, momentum and energy on the basis of the relativistic velocity transformation.[6,7,8] The inertial reference frames involved are K(XOY) and K'(X'O'Y'). The corresponding axes of the two frames are parallel to each other and the OX(O'X') axes overlap. K' moves with constant velocity V relative to K' in the positive direction of the overlapped axes. At the common origin of time in the two frames (t=t'=0) the origins of the two frames are instantaneously locate at the same point in space.

## 2. Weighing a moving body: rest mass and relativistic mass

We describe first a thought experiment involving a balance set in a gravity field and with its central pillar bound to the Earth (the K reference frame).[9] The pans move with constant velocity $V_0$ in the positive and in the negative direction of the OX axis, respectively, starting at a time $t=0$ from the central pillar. After a time $t$ the lengths of the beams are equal to each other and equal to $V_0 t$. We place two identical bodies **1** and **2** in the left and in the right pans of the balance, respectively. Under such conditions, the balance is in a state of equilibrium with its central pillar parallel to the lines of gravity and with its beams perpendicular to them (Figure 1a). In accordance with the first postulate the balance should be in a state of equilibrium when observed from the K' reference frame as well, which has its origin O' in the left pan of the balance. In the K' reference frame, the rest frame of the left pan, located at its origin O' (Figure 1b). Relative to this frame the central pillar moves with velocity $V_0$ whereas the right pan moves with velocity $V$ both in the positive direction of the overlapped axes. In accordance with the relativistic velocity transformation we have

$$V = \frac{2V_0}{1+\frac{V_0^2}{c^2}} \qquad (1)$$

resulting that in the K' reference frame the lengths of the two beams are no longer equal to each other. The left beam has a length $L_1' = V_0 \Delta t_0'$ (2) whereas the length of the right beam is $L_2' = (V-V_0)\Delta t_0'$ (3), $\Delta t_0' = t_0' - 0$ representing the time duration of the motion measured in the rest frame of the left pan.

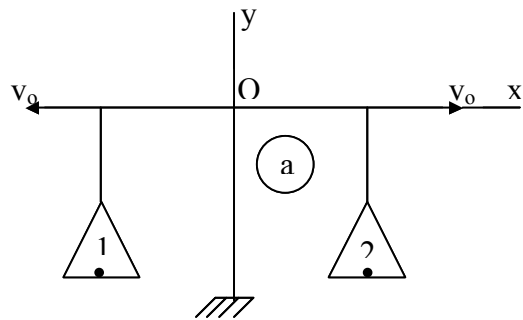

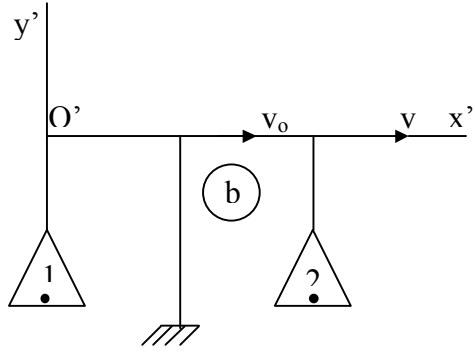

*Figure 1 a. The balance as detected from the K frame, the rest frame of the central pillar. b. The balance as detected from the K' frame, the rest frame of the left pan.*

Solving Eq.(1) for $V_0$ we obtain

$$V_0 = \frac{c^2}{V}\left[1 - \sqrt{1 - \frac{V^2}{c^2}}\right] \tag{4}$$

and so

$$\frac{L_1'}{L_2'} = \frac{1}{\sqrt{1 - \frac{V^2}{c^2}}} \tag{5}$$

The body in the left pan is in a state of rest in the K' frame and relativists say that observers of that frame measure its rest mass $m_0$. The body in the right pan moves with velocity $V$ relative to the K' frame, observers of that frame measuring its relativistic mass. The balance is in a state of equilibrium in the K' frame if

$$m_0 L_1' = m L_2' \tag{6}$$

so

$$m = \frac{m_0}{\sqrt{1 - \frac{V^2}{c^2}}} \tag{7}$$

In the experiment presented above observers measure the gravitational mass. Changes in the scenario show that Eq. (7) holds in the case of inertial mass as well. Real experiments confirm its validity.[10]

Eq. (7) establishes a relationship between two physical quantities measured by observers of the same frame, $m_0$ when the particle is in a state of rest and $m$ when it moves with velocity V.

**2. Relativistic mass and relativistic momentum: How do they transform?**

The scenario we follow involves a particle moving with velocity $\mathbf{u}(u_x, u_y)$ relative to K and with velocity $\mathbf{u}'(u_x', u_y')$. It moves in the XOY (X'O'Y') plane. At the common origin of time (t = t'= 0) the particle is instantaneously located at the point where the origins O and O' meet each other. Detected from K the particle moves along a direction θ whereas detected from K' it moves along a direction θ', relative to the positive direction of the overlapped OX(O'X') axes. Our derivation does not involve other scenarios.

The relativistic velocity transformation tells us that **u** and **u'** and their components add as

$$u_x = \frac{u'_x + V}{1 + \frac{Vu'_x}{c^2}} = u' \frac{\cos\theta' + \frac{V}{u'}}{1 + \frac{Vu'}{c^2}\cos\theta'} \tag{8}$$

$$u_y = \frac{\sqrt{1 - \frac{V^2}{c^2}}\, u'_y}{1 + \frac{Vu'_x}{c^2}} = u' \frac{\sqrt{1 - \frac{V^2}{c^2}}\sin\theta'}{1 + \frac{Vu'}{c^2}\cos\theta'} \tag{9}$$

$$u = u' \frac{\sqrt{(\cos\theta' + \frac{V}{u'})^2 + (1 - \frac{V^2}{c^2})\sin^2\theta'}}{1 + \frac{Vu'}{c^2}\cos\theta'}\ . \tag{10}$$

The angles θ and θ' transform as

$$\cos\theta = \frac{u_x}{u} = \frac{\cos\theta' + \frac{V}{u'}}{\sqrt{(\cos\theta' + \frac{V}{u'})^2 + (1 - \frac{V^2}{c^2})\sin^2\theta'}} \tag{11}$$

$$\sin\theta = \frac{u_y}{u} = \frac{\sqrt{1 - \frac{V^2}{c^2}}\sin\theta'}{1 + \frac{Vu'}{c^2}\cos\theta'} \tag{12}$$

Observers from K will consider that the momentum of the particle is

$$\mathbf{p} = m\mathbf{u} \tag{13}$$

whereas observers at rest in K' will consider that the momentum of the same particle is

$$\mathbf{p}' = m'\mathbf{u}'. \tag{14}$$

Many authors start with similar assumptions but use the conservation of momentum law. We avoid such an approach.

The thought experiment presented in Section 2 warns the two physicists that *m* and *m'* are relativistic masses and they know that **u** and **u'** are relativistic velocities.

Combining equations (13) and (14) and taking into account Eq. (10) we obtain

$$\frac{p}{m} = \frac{p'}{m'} \cdot \frac{\sqrt{(\cos\theta' + \frac{V}{u'})^2 + (1 - \frac{V^2}{c^2})\sin^2\theta'}}{1 + \frac{Vu'}{c^2}\cos\theta'} \quad . \tag{15}$$

Eq. (15) suggests considering that the magnitude of the momentum transforms as

$$p = f(V) p' \sqrt{(\cos\theta' + \frac{V}{u'})^2 + (1 - \frac{V^2}{c^2})\sin^2\theta'} \tag{16}$$

mass transforming as

$$m = f(V) m' (1 + \frac{Vu'}{c^2}\cos\theta'). \tag{17}$$

The unknown function *f(V)* depends only on the relative velocity *V* and not on the physical quantities involved in the transformation process (to one particle in the K frame corresponds a single particle in K' and vice-versa, a condition called "linearity of the transformation equations"). Physicist *R* considers the situation when the particle is at rest in K' (*u'=0*) moving with velocity *V* relative to himself. Under such conditions observers from K' measure the rest mass $m_0$ of the involved particle, and Eq. (17) becomes

$$m = f(V) m_0. \tag{18}$$

Comparing Eqs. (7) and (18) we obtain

$$f(V) = \frac{1}{\sqrt{1 - \frac{V^2}{c^2}}} = \gamma(V). \tag{19}$$

The components of the momentum transform as

$$p_x = p\cos\theta = \gamma(V) p'(\cos\theta' + \frac{V}{u'}) = \gamma(V)(p'_x + Vm') \tag{20}$$

$$p_y = p\sin\theta = p'\sin\theta' = p'_y. \tag{21}$$

As we see, the OY(O'Y') component of the momentum is a relativistic invariant. We can present now Eq. (17) as

$$m = \gamma(V) m' (1 + \frac{Vu'}{c^2}\cos\theta') = \gamma(V)(m' + \frac{V}{c^2} p'_x) \tag{22}$$

Relativists call **p** and **p'** defined by Eqs. (13) and (14) relativistic momentum, and we can express them as

$$\mathbf{p} = \frac{m_0 \mathbf{u}}{\sqrt{1 - \frac{u^2}{c^2}}} \tag{23}$$

in K and

$$\mathbf{p}' = \frac{m_0 \mathbf{u}'}{\sqrt{1 - \frac{u'^2}{c^2}}} \tag{24}$$

in K'.

For those physicists who ban the concept of relativistic mass[11,12,13] because it often leads to confusion, we derive transformation equations free of this concept. We mention that the concept of relativistic mass has its defenders.[14]

The concept of relativistic momentum and its relation to relativistic mass play a fundamental part in relativistic dynamics. In most introductory treatments, the expression for the relativistic momentum is found by considering collisions between two particles from two inertial reference frames in relative motion using the law of momentum conservation.[15,16,17] We underline that we have derived it without using conservation laws or collisions.

For more fun, we present a two-line derivation of the relativistic momentum. Let R' be an observer commoving with the particle. An observer R from K could define the particle's momentum as

$$p = m_0 \frac{\Delta r}{\Delta \tau} \qquad (25)$$

and its instantaneous position using the Cartesian space coordinates ($x, y, r = \sqrt{x^2 + y^2}$). In Eq. (25) $\Delta r$ represents the change in the length of the position vector in K during the proper time interval $\Delta \tau$ measured by R'. Measured by clocks in the K frame the trip of the particle lasts a time interval $\Delta t$ related to $\Delta \tau$ by the time dilation formula

$$\Delta \tau = \sqrt{1 - \frac{u^2}{c^2}} \Delta t \qquad (26)$$

which replaced in Eq. (25) leads to the expression for the relativistic momentum given by Eq.(23)

**3. The photon**

We replace the particle we have considered so far with a photon, "a quantum of light energy, regarded as a discrete particle, having zero rest mass, no electric charge and an indefinitely long lifetime" and moves with velocity c relative to all inertial reference frames. An electromagnetic wave carries energy $E$ (heats a body it hits) and momentum p (exerts pressure on the body it hits).[15] Equation

$$p = \frac{E}{c}, \qquad (27)$$

relates momentum and energy the electromagnetic wave carries. Considering that an invariant number of photons carry $p$ and $E$ then equation

$$p_c = \frac{E_c}{c} \qquad (28)$$

relates momentum and energy carried by a single photon. Detected from K the photon moves along a direction $\theta_c$ whereas detected from K' it moves along a direction $\theta_c'$ relative to the positive direction of the overlapped axes, respectively. The components of its velocity are

$$c_x = c \cos \theta_c \qquad (29)$$
$$c_y = c \sin \theta_c \qquad (30)$$
$$c_x' = c \sin \theta_c' \qquad (31)$$
$$c_y' = c \sin \theta_c'. \qquad (32)$$

Eqs. (11) and (12) tell us that the angles $\theta$ and $\theta'$ transform as

$$\cos \theta_c = \frac{\cos \theta_c' + \frac{V}{c}}{1 + \frac{V}{c} \cos \theta_c'} \qquad (33)$$

$$\sin\theta_c = \frac{\sqrt{1-\frac{V^2}{c^2}}\sin\theta'_c}{1+\frac{V}{c}\cos\theta'_c} \quad (34)$$

accounting for the aberration of light effect.[18]

The OX(O'X') component of the photon's momentum transforms as

$$p_{x,c} = \gamma(V)p'_{x,c}(\cos\theta'_c + \frac{V}{c}) \quad (35)$$

whereas its OY(O'Y') component is a relativistic invariant

$$p_{y,c} = p'_{y,c}. \quad (36)$$

The magnitude of the photon's momentum transforms as

$$p_c = \gamma(V)p'_c(1+\frac{V}{c}\cos\theta'_c). \quad (37)$$

Because the quotient $\frac{E}{p} = c$ is a relativistic invariant, the photon's energy should transform as

$$E_c = \gamma(V)E'_c(1+\frac{V}{c}\cos\theta'_c). \quad (38)$$

Multiplying both sides of Eq. (22) with $c^2$, it becomes

$$mc^2 = \gamma(V)m'c^2(1+\frac{Vu'}{c^2}\cos\theta'). \quad (39)$$

Comparing Eq. (20) with Eq. (38) and Eq. (21) with Eq. (36), respectively, we see that we can obtain the transformations for $p_{x,c}$ and $p'_{x,c}$ from the corresponding transformations for $p_x$ and $p'_x$ by simply replacing $u'$ by $c$. Eq. (39) leads to Eq. (38) if we follow the same procedure as in the case of the momentum ($u' \to c$) considering that

$$E = mc^2 = \frac{m_0 c^2}{\sqrt{1-\frac{u^2}{c^2}}} \quad (40)$$

and

$$E' = m'c^2 = \frac{m_0 c^2}{\sqrt{1-\frac{u'^2}{c^2}}} \quad (41)$$

represent the relativistic energy of the tardyon in K and in K' respectively. With the new notations Eq. (39) becomes

$$E = \gamma(V)E'(1+\frac{Vu'}{c^2}\cos\theta'). \quad (42)$$

We stress that $mc^2$ and $m'c^2$ have the physical dimensions of energy.

Relativists call $E(E')$ relativistic energy. If the particle is at rest in K' ($u'=0$) observers in that frame measure its rest energy $E_0$ related to $E$ by

$$E = \gamma(V)E_0. \quad (43)$$

Because the single supplementary energy our free particle could posses is its kinetic energy $E_k$ we obtain for it

$$E_k = E - E_0 = m_0 c^2 \left(\frac{1}{\sqrt{1-\frac{V^2}{c^2}}} - 1\right) = E_0[\gamma(V) - 1]. \tag{44}$$

We have derived the expression for the kinetic energy without using the work energy theorem.[17]

Experiment proves the validity of Eq. (44).[18]

We now have all the elements needed to eliminate the concept of relativistic mass. We can express the OX(O'X') component of the momentum as a function of energy as

$$p_x = \gamma(V)(p_x' + \frac{V}{c^2} E') \tag{45}$$

and the energy as a function of the O'X' component of the momentum as

$$E = \gamma(V)(E' + V p_x'). \tag{46}$$

Eqs. (41) and (42) express the so called mass-energy equivalence in K and in K' respectively. They tell us that we can characterize the inertia of a tardyon by its mass or by its energy. Presented as

$$\Delta E = c^2 \Delta m \tag{47}$$

Eq. (47) tells us that an increase in the energy of a particle increases its inertia.

**4. An identity that generates the energy-momentum four vector**

We start with the obvious identity

$$\frac{1-\frac{u^2}{c^2}}{1-\frac{u^2}{c^2}} = 1. \tag{48}$$

The relativistic identity still holds if we multiply both sides of Eq. (48) with the invariant $m_0^2 c^4$. Using the notations have introduced above, it leads to

$$\frac{E^2}{c^2} - p^2 = m_0^2 c^4 \tag{49}$$

and we have derived that way a relativistic invariant. In the general case it becomes

$$\frac{E^2}{c^2} - p_x^2 - p_y^2 - p_z^2 = m_0^2 c^2 = invariant. \tag{50}$$

Relativists introduce the concept of energy-momentum three vector, the components of which are $(p_x, p_y, p_z, \frac{E}{c})$ in K and $(p_x', p_y', p_z', \frac{E'}{c})$ in K' respectively.

In accordance with Eq. (50) we have

$$\frac{E^2}{c^2} - p_x^2 - p_y^2 - p_z^2 = \frac{E'^2}{c^2} - p_x'^2 - p_y'^2 - p_z'^2 = m_0^2 c^2 = invariant \tag{51}$$

in a three space dimensions the OZ(O'Z') component of the momentum being a relativistic invariant.

Applied to the case of a photon ($p_c = \frac{E_c}{c}$), Eq. (49) tells us that its rest mass is equal to zero, a photon existing only when it moves with velocity c. The energy of a photon incident on a blackbody increases its internal energy and in accordance with the postulate of mass-energy equivalence (Eq. (47)) it increases its inertia as well.

The fact that the energy of a photon and the frequency $\nu$ of the electromagnetic oscillations that take place in the electromagnetic wave are related by[19]

$$E_c = h\nu \tag{52}$$

in K and

$$E_c' = h\nu' \tag{53}$$

in K' where $h$ represents Planck's constant, implies that the frequency of the electromagnetic oscillations transforms as

$$\nu = \gamma(V)\nu'(1 + \frac{V}{c}\cos\theta'). \tag{54}$$

Eq. (54) accounts for the Doppler Effect.[20]

**4. Conclusions**

We derive the fundamental equations of relativistic dynamics starting with a thought experiment, which warns physicists that we should make a net distinction between the mass of a particle in a state of rest (invariant or rest mass) and the mass of the same particle when it moves relative to the observer (relativistic mass). The interpretation of a thought experiment in which an observer weights a moving body involves the first postulate and the transformation of relativistic velocity leading to a relationship between rest mass and relativistic mass.

We also derive the expressions for the relativistic momentum, for the transformation for mass and momentum without using conservation laws of momentum and mass.

The concept of relativistic energy is introduced considering that special relativity ensures a smooth transition from the formulas that describe the dynamic properties of a tardyon to the formulas that describe the dynamic properties of a photon ('Natura non facit saltus" Leibniz) and the transformation of energy is derived without using the laws of momentum and energy conservation.

For those who want to avoid the concept of relativistic mass we have derived formulas that ban it.

Our approach reveals the importance of thought experiments in simplifying the teaching of special relativity and makes the relativistic effects less counterintuitive. It also reveals the kinematical roots of relativistic dynamics.

Deriving all the fundamental equations of relativistic dynamics from a single scenario makes relativistic dynamics easy to teach and to learn.

**References**
[1]Asher Peres, "Relativistic telemetry," Am.J.Phys. **55**, 516-519 (1987)
[2]Margaret Stautberg Greenwood, "Relativistic addition of velocities using Lorentz contraction and time dilation," Am.J.Phys. **50** 1156-1157 (1982)
[3]N. David Mermin, "Relativistic addition of velocities directly from the constancy of the velocity of light," Am.J.Phys. **52,** 1119-1124 (1984)
[4]L. Sartori, "Elementary derivation of relativistic velocity addition law," Am.J.Phys. **63**, 81-82 (1995)
[5]W.N. Mathews Jr. "Relativistic velocity and acceleration transformation," Am.J.Phys. **73**, 45-51 (2005)
[6]Bernhard Rothenstein and George Eckstein, "Lorentz transformation directly from the speed of light," Am.J.Phys. **63**, 1150 (1995)
[7]Bernhard Rothenstein and Aldo Desabata, "An uncommon way to special relativity," Eur.J.Phys. **18**, 263-266 (1997)
[8]Edward Kapuscik, "Comment on Lorentz transformations directly from the speed of light, by B.Rothenstein and G.Eckstein [Am.J.Phys. **63**, 1150 (1995)]" Am.J.Phys. **65**, 1210 (1997)

[9] Ling Tsai, "The relation between gravitational mass, inertial mass and velocity," Am.J.Phys. **54,** 340-342 (1956)

[10] W.G.V. Rosser, *An Introduction to the Theory of Relativity,* (Butherworth, London,1964) pp. 193-199

[11] L.B.Okun, "The concept of mass," Phys. Today, **42**, 31-36 (1989)

[12] Carl G. Adler, "Does mass depend on velocity, dad?", Am.J.Phys. **55,** 739-743 (1987)

[13] Robert W. Brehme, "The advantage of teaching relativity with four vectors," Am.J.Phys. **36**, 896-901 (1968)

[14] T.R. Sandin, "In the defense of relativistic mass," Am.J.Phys. **59**, 1032-1036 (1981)

[15] P.C.Peters, An alternate derivation of relativistic momentum," Am.J.Phys. **54**, 804-808 (1986)

[16] C.Moller, *TheTheory of Relativity,* (Clarendon Press, Oxford, 1968) Ch.3

[17] G.W. Lewis and R.C. Tolman, Philos. Mag. **18**, 510 (1909). The authors start with the same assumptions as we do but use the momentum conservation.

[18] Robert Resnick, *Introduction to Special Relativity,* (John Wiley and Sons, Inc. New York, 1968) pp.84-87